# Schmitt-Trigger-based Recycling Sensor and Robust and High-Quality PUFs for Counterfeit IC Detection

*Cheng-Wei Lin, Jae-Won Jang and Swaroop Ghosh*
Department of Computer Science and Engineering
University of South Florida
Tampa, FL, USA, 33620
{chengwei, jjang3, sghosh }@mail.usf.edu

**Abstract:** *We propose Schmitt-Trigger (ST) based recycling sensor that are tailored to amplify the aging mechanisms and detect fine grained recycling (minutes to seconds). We exploit the susceptibility of ST to process variations to realize high-quality arbiter PUF. Conventional SRAM PUF suffer from environmental fluctuation-induced bit flipping. We propose 8T SRAM PUF with a back-to-back PMOS latch to improve robustness by 4X. We also propose a low-power 7T SRAM with embedded Magnetic Tunnel Junction (MTJ) devices to enhance the robustness (2.3X to 20X).*

## Introduction and Motivation

Chip recycling involves scavenging and reusing the aged but functionally correct ICs in new systems. Although the machines might operate correctly, the operating speed and energy-efficiency will be degraded due to prior usage. Detecting recycled ICs are essential to improve the security and trustworthiness of the integrated systems [1]. Conventional techniques exploit the temporal degradation of circuit performance to isolate the recycled ICs. These techniques rely on aging mechanisms such as Bias Temperature Instability (BTI) and Hot-Carrier Injection to degrade the Ring Oscillator (RO) based sensor circuit. The performance degradation of degraded RO is compared with the fresh RO to identify the recycled ICs.

One of the primary challenges in recycling detection is the process variation between the aged (or stressed) and fresh RO at t=0. As illustrated in Fig. 1(a), the distribution of delay difference between fresh and stressed RO at t=0 under process variations, as well as the shift of distribution after different amount of stress periods. It is evident that the chips on the left will be falsely pronounced as recycled and chips on right will mask fine grained recycling. Isolating the recycled ICs from the genuine ones for arbitrarily small amount of usage (few seconds to minutes) is a challenging task.

The combating die recovery (CDR) sensor [2] employs two ROs, a reference and a stressed RO. Although effective, the proposed methodology can only identify chips that are used for at least few months. Therefore, few hours or days of usage will go undetected. The Anti-Fuse (AF) based design [2] consists of a counter and an AF ROM block. The usage period of the chip is determined based on the number of cells that have been flipped. Although the AF based sensor can detect fine usage interval, the accuracy is restricted by the size of the AF ROM.

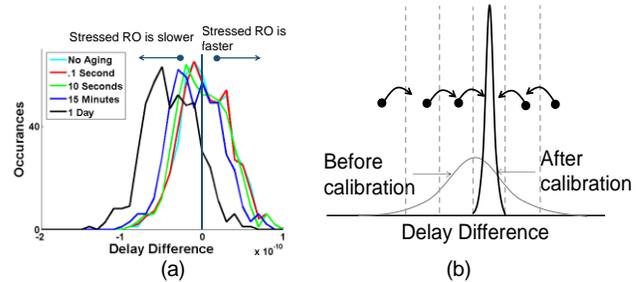

Fig. 1 (a) Simulation results showing distribution of delay difference between fresh and stressed RO at t=0 and, movement of distribution with various amount of recycling; (b) Compensating process variations by self-calibration. The faster RO is selected as the "stressed" RO and slowed down through addition of extra programmable loads. Since delay difference is calculated as "fresh" RO delay minus "stressed" RO delay, the resulting distribution will always be positive and tighter after calibration.

IC cloning or counterfeiting involves copying the designs and fabricating them with the intention to impersonate the authentic chip, gain access to secure content and/or leak secret information [3]. Physically Unclonable Function (PUF) [4-11] is the prime ingredient to prevent cloning. It replaces the hard-coded key in the IC with specifically designed circuits that work on the principle of challenge-response. The response to a particular challenge is based on the physical properties of the chip (e.g., process). The unclonability of the PUF makes the response hard to predict by the adversaries. Several types of PUFs have been proposed [4-11] however due to its simplicity, SRAM PUFs [8] and arbiter-PUFs [9] are a widely accepted designs. One of the key PUF design challenge is to ensure repeatability of the response under environmental fluctuations and consecutive access. Typical metric used to characterize the PUF quality is inter and intra-die Hamming Distance (HD) [6].

*Proposed Solutions:* Traditionally, precautions are taken to suppress the aging effects such as Negative Bias Temperature Instability (NBTI), Positive Bias Temperature Instability (PBTI) and Hot Carrier Injection (HCI) [12-13] in circuits. This work exploits the aging mechanisms to enhance the sensitivity of recycling sensors to usage interval. The proposed sensors amplify the effect of NBTI and PBTI through voltage boosting and under-drive and, HCI by using Schmitt Trigger [13]. Our analysis indicate that the proposed sensors can reliably detect IC recycling of few seconds. *To the best of our knowledge this is the first attempt to design recycling sensor that leverages the three types of transistor aging mechanisms for chip authentication.* We observe that

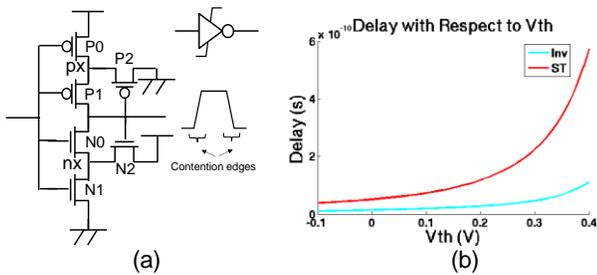

Fig. 2 (a) Schmitt-trigger schematic; (b) Sensitivity of ST delay with respect to added $V_{TH}$ of the transistors ($\Delta V_{TH}$ of all transistors are swept together to mimic aging). ST shows as much as 5X higher delay sensitivity than inverter.

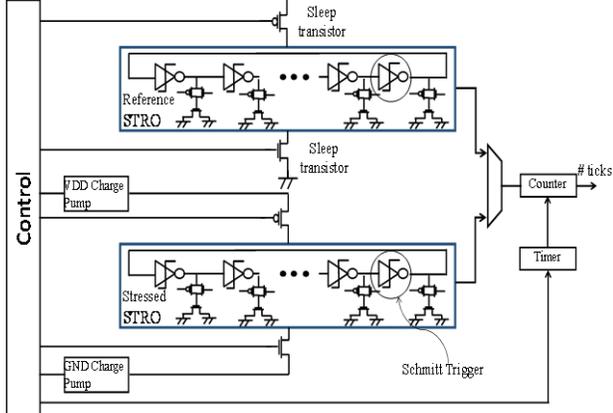

Fig. 3 Schematic of STRO based recycling sensor. The reference STRO is slept during normal mode to save power and disable aging. Therefore, reference STRO always provides oscillation corresponding to brand new IC. The stressed STRO receives VDD (GND) from positive (negative) charge pumps. A 4-phase positive and negative charge pumps are used to generate a low ripple VDD and GND supply. The charge pumps are turned-off after stressing period is over. During the sensing, the stressed STRO operating at nominal VDD and GND provide degraded oscillations. A counter counts the number of ticks of reference and stressed RO with respect to a fixed timer interval. The difference between ticks represents the frequency difference between fresh and degraded oscillator i.e., the usage interval.

ST is inherently more prone to process variation while being robust to other sources of noise. Therefore we extend the application of ST to realize novel arbiter-based PUF. Our analysis indicate better inter and intra-die HD for ST-PUF compared to conventional inverter based PUF under process variations and environmental fluctuations. We also propose low-overhead 8T and non-volatile (NV) 7T SRAM PUF to enhance the stability and repeatability of conventional SRAM PUF. To best of our knowledge, this is the first attempt to improve the stability of SRAM PUF using 8T bitcell and a MTJ. Each of the proposed techniques are described in detail below.

## Schmitt-Trigger-based Recycling Sensor

In this section, we describe the operation of ST and the proposed self-calibrating recycling sensor operating at boosted voltage.

*Overview of ST:* ST is widely used to perform robust computation in presence of noisy inputs (such as IO pads). Application of ST for low voltage circuits such as subthreshold SRAM has also been proposed. ST modulates the switching threshold of the inverter based on the direction of input transition. The schematic of a conventional ST is shown in Fig. 2(a). P0-1 and N0-1 are stacked transistors that form an inverter configuration whereas P2 and N2 are feedback transistor that create contention at the output. When the output is at '1' and input toggles from 0→1, N2 prevents node nx to discharge the output effectively moving the switching threshold of inverter high. Similarly, P2 prevents node px to charge the output even during 1→0 input transition moving the switching threshold of inverter low.

*Sensitivity of ST to Process Variation:* Although ST is immune to input noise, its delay is sensitive to the strength of feedback transistors P2 and N2 (and stacked transistors P0, P1 and N0, N1). Aging in these transistors is amplified in terms of delay. Fig. 2(b) compares inverter and ST delay sensitivity with respect to aging in 22nm predictive technology model [15]. For this simulation $\Delta V_{TH}$ is added to the $V_{TH}$ of all transistors to mimic aging. The value of $\Delta V_{TH}$ is swept to observe the sensitivity of inverter and ST to different amount of aging. It is evident that ST is as much as 5X more sensitive to aging induced $V_{TH}$ shift than inverter. Based on the above observation, we propose ST-based recycling sensor. *Note that susceptibility of ST delay on $V_{TH}$ shift also makes it more sensitive to process variations. However, we nullify the effect using self-calibration during initialization.* This allows us to exploit the variability of ST only due to aging for recycling sensing.

*Sensor Design:* We propose a combination of four techniques to address the challenges above mentioned: (a) employing ST instead of inverter in the RO to amplify the impact of aging on delay difference distribution, (b) using self-calibration to reduce the spread of delay difference in ST, as conceptually illustrated in (Fig. 1(b)). (c) boosting VDD and under-driving GND using light-weight charge pumps [16] to accelerate BTI (NBTI in PMOS and PBTI in NMOS) effect and (d) employing high-$V_{TH}$ devices to further amplify the impact of aging on delay. The objective is to tighten the delay difference spread before stress and create a large shift in delay difference distribution even for fine grained usage interval to allow a clear distinction between new and recycled chips. The impact of HCI [17-18] is enhanced naturally by ST due to larger slew rate of output when the input is making a transition.

Fig. 3 shows the schematic of Schmitt Trigger RO (STRO) based recycling sensor. Extra loading capacitors are added in the internal nodes of the fresh and stress RO. The capacitors on the fast RO are selected based on the control signals during calibration phase to equalize the delay of both ROs.

*Simulation Setup:* For analysis, we model intra-die process variations in 22nm predictive technology [15]. The process variation is modeling by lumping the variations in transistor threshold voltage ($V_{TH}$). The mean and standard deviation of intra-die $V_{TH}$ shift is assumed to be (0, 50mV). The actual $V_{TH}$ of the transistor is the summation of $V_{TH}$ (nominal) and $V_{TH}$ (intra). The simulation temperature is fixed at 298K. For high-

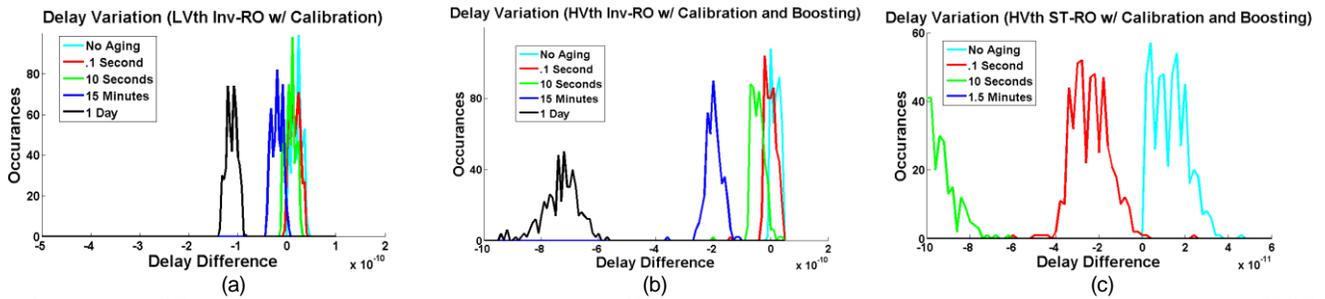

Fig. 4 (a) Inverter RO recycling sensor with calibration, (b) inverter RO recycling sensor with high-$V_{TH}$ and calibration, and (c) high-$V_{TH}$ STRO recycling sensor with calibration and boosting.

$V_{TH}$ devices, we added 300mV on top of nominal $V_{TH}$. For all simulations, we employed 31-stage RO. A total of 500 Monte Carlo points have been simulated to observe the impact of variations. For VDD boosting we used a simple RO-based 4-phase positive and negative charge pumps to generate a low ripple VDD and GND supply. The positive (negative) charge pumps provide 1.4V (-0.25V) from 1V of supply voltage. The peak-to-peak ripple is less than 50mV (omitted for brevity).

We design and evaluate the effectiveness of each design technique separately. Recycling of 0.1s, 10s, 1.5min, 15min and 1 day is simulated to determine the detection ability. Reliability analysis is performed using MOSRA [19] to measure the impact of both HCI and BTI degradations.

*Simulation Results:* Fig. 4(a)-(b) shows the distributions for inverter RO sensor. From Fig. 1(a) it is evident that even 1 day of recycling cannot be detected due to large process-variation induced overlap with original distribution. Self-calibration enables detection of 1 day of usage reliably and 15min of usage with <5% error. Combining high-$V_{TH}$ and self-calibration shows detection of 15min of usage with negligible error (<1%) and 10s of usage with ~10% error. These results underscore the impact of self-calibration, high-$V_{TH}$ and boosting in improving the recycling detection ability by several orders of magnitude. Fig. 4(c) plots the distributions for inverter STRO sensor. Combining high-$V_{TH}$, self-calibration and addition of VDD and GND boosting successfully separates the distributions for clear identification of >10s of usage and reduces the error in 0.1s of usage detection to <1%. Considering the original detection accuracy of 1 day, the proposed sensor is $8.64 \times 10^5$X more accurate.

### ST-based Arbiter PUF

The conventional arbiter PUF [20] contains an arbiter and two identical delay paths. For authentication, a signal is raced in two delay paths (the exact path is determined by the challenge). The arbiter outputs response (0 or 1) depending on early arriving path. The scheme employs delay difference to minimize environmental fluctuation induced errors (i.e. temperature and voltage variation). Thus, the output of the arbiter PUF is largely depend on the delay variation of the two delay paths. The conventional design often produce error response for small delay variation. Here, we propose a novel arbiter-based PUF with ST as the building block (Fig. 6(a)) for the delay path. From Fig. 6(a), ST experiences larger delay variation allowing clear identification of the winner of the race. Hence the reliability of the ST PUF is ensured under severe environmental fluctuations.

Fig. 5(a) shows the delay difference distribution for 4-stage inverter and ST-PUF for all sets of challenges. It can be observed that ST-PUF provides wider spread of distribution. The plot also indicates balanced 0 and 1 response. The corresponding plot for 10 and 20-stage PUFs are shown in Fig. 5(b). The ST-PUF show 44% improvement in mean and 10% in sigma for the intra-die HD. This indicates that the proposed ST-PUF not only makes the design more stable it also tightens the outlier cases. We executed the NIST benchmarks on both PUF designs. The ST-PUF improves 5 out of 9 benchmarks outlining the improved stability and quality.

### 8T and NV 7T SRAM PUF

In the conventional 6T SRAM PUF (Fig. 6(b)), the initial state of bitcell after power-ON is purely random due to process variations. The address of the array is used as a challenge, and the data is used as response. Once initialized, the content of the SRAM is expected to be constant. However, the environmental fluctuations can cause bit flipping. In addition, the PUF response is expected to be constant for subsequent initialization.

We propose an 8T SRAM PUF with embedded latch to stabilize the bit under environmental fluctuations. By keeping the latch ON it is possible to maintain the bitcell content for repetitive power-ON sequence. Although robust, the 8T SRAM PUF consume static power. We also propose a NV 7T SRAM with embedded MTJ devices to enhance the robustness while maintaining device leakage. Note that the functioning and structure of the proposed NV SRAM PUF is different than the NV SRAM proposed for cache [21-23].

During the powering sequence of 8T PUF, the signal LE is kept high to disable the latch (Fig. 6(c)). Once the bits are

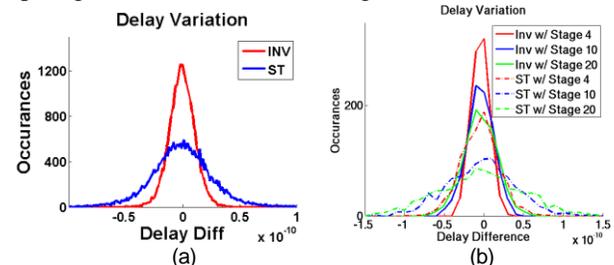

Fig. 5 Delay difference distribution of arbiter-PUF and ST-PUF. (a) 4-stage PUFs and (b) 4, 10 and 20 stages. ST-PUF shows wider spread of distribution.

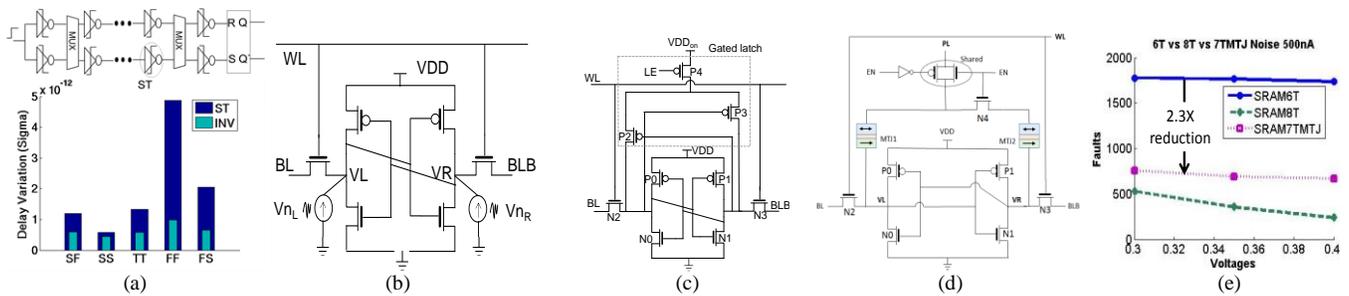

Figure 6 (a) ST arbiter PUF and simulation showing ST is more prone to process variations compared to inverter; (b) 6T-SRAM PUF with noise; (c) 8T-SRAM PUF; (d) 7T NV SRAM PUF; (e) Faults vs Voltages for designs initialized, LE is pulled down to turn ON the sleep transistor and enable the latch. The latch reinforces the bitcell content depending on the internal node voltage. The node holding a '1' is strengthened by the addition of extra PMOS transistor. *This is functionally equivalent to upsizing the PMOS dynamically*. Note that since the internal node is strongly held than before, the 8T PUF is robust to voltage and temperature variations. The latch could be kept active while the SRAM is slept or powered down. This is beneficial for repeatability because the latch will reinforce the cell content when the SRAM is powered-ON again.

The NV 7T SRAM PUF (Fig. 6(d)) includes one access transistor (N4) and two MTJs (MTJ1 and MTJ2) in the internal nodes for permanent memorization of the values after registration. The MTJs are gated through a shared mux. The purpose of the mux is two-fold: (a) isolate the MTJ random initialization from the random initialization of SRAM. Once the SRAM is initialized the values are written to the MTJs by enabling the mux and, (b) reduce the leakage through the MTJs by disabling them when not needed. Note that the gating mux and additional access transistor is designed using high-$V_{TH}$ to reduce the leakage.

After the power-ON sequence the SRAM is initialized to random values. The values are programmed in the MTJs by turning ON the enable signal (EN) and access transistor N4 and following two steps: (a) When PL=0, the node storing a '1' writes a high resistance to the MTJ. (b) When PL=1, the node storing '0' writes a low resistance to the MTJ.

Once the MTJs are programmed, it remains there and reinforces the values. The node with '0' value and low MTJ resistance provides additional strength to the corresponding NMOS transistor. After programming the MTJs, the SRAM contents could be read (registration). When the power is turned-OFF and turned back ON, the MTJ with low resistance reinforces the '0' side ensuring that the SRAM is brought to the same state as before. This is achieved by enabling the gating multiplexer and N4 ON with PL=0 before the next initialization.

The robustness of 6T, 8T and NV 7T PUFs are compared in Fig. 6(e). The proposed NV 7T PUF under repetitive powering, it shows 2.3X better robustness compared to 6T PUF.

**Acknowledgement**

This paper is based on work supported by Semiconductor Research Corporation (#2442.001) and NSF CNS #1441757.